\documentclass{PoS}

\title{Pointing the James Webb Space Telescope through lensing clusters - can the first stars and galaxies be detected?}

\ShortTitle{Pointing the James Webb Space Telescope through lensing clusters}

\author{\speaker{Erik Zackrisson}\\
        Department of Astronomy, Stockholm University\\
        E-mail: \email{ez@astro.su.se}}

\abstract{The James Webb Space Telescope (JWST), scheduled for launch in 2014, is expected to revolutionize our understanding of the high-redshift Universe. Even so, many of the most interesting sources that may be hiding at redshifts $z\gtrsim 10$ (population III stars, dark stars, population III galaxies) are likely to be intrinsically too faint for JWST. Here, we explore the prospects of searching for the first stars and galaxies by pointing JWST through foreground lensing clusters. Observations of this kind can reach significantly deeper than the currently planned JWST ultra deep field in just a fraction of the exposure time, but at the expense of probing a much smaller volume of the high-redshift Universe. We also present {\it Yggdrasil}, a spectral synthesis code for modelling the first galaxies, and use it to derive the masses of the faintest pop I, II and III galaxies that can be detected through broadband imaging in JWST ultra deep fields.}

\FullConference{Cosmic Radiation Fields: Sources in the early Universe\\
		 November 9-12, 2010\\
		 Desy, Germany}

\begin{document}

\section{Introduction}
The first stars forming in the history of the Universe are generically predicted to be very massive ($\gtrsim 10$--$100\ M_\odot$; e.g. \cite{Bromm et al. a,Nakamura & Umemura,Greif & Bromm,Stacy et al.,Clark et al.}) -- owing both to the lack of efficient coolants in the almost metal-free gas out of which these objects formed, and the higher temperature of the cosmic microwave background radiation at high redshifts. These population III objects (herafter pop III) are expected to start forming in isolation or in small numbers within $\sim$10$^5$--$10^6 M_\odot$ dark matter halos (so-called minihalos) at redshifts $z\approx 20$--50 (e.g. \cite{Stacy et al.,Clark et al.,Tegmark et al.,Yoshida et al.}), but the prospects of detecting such stars on an individual basis appear bleak (e.g. \cite{Gardner et al.,Greif et al.,Rydberg et al.}), at least before they go supernovae \cite{Weinmann & Lilly,Whalen & Fryer}. 

However, pop III stars may also continue to form within the more massive halos ($\gtrsim 10^7-10^8 \ M_\odot$) hosting some of the earliest galaxies\footnote{Due to the low stellar population masses of these objects, they are sometimes also -- and perhaps more appropriately -- referred to as pop III star clusters} at $z\lesssim 15$ \cite{Scannapieco et al.,Tornatore et al.,Johnson et al. a,Johnson et al. b,Stiavelli & Trenti,Salvaterra et al.,Johnson}. Since the {\it very first} galaxies are predicted to form in high-density regions that have been pre-enriched by pop III stars in minihalos, these systems are not expected to be metal-free, and are most likely dominated by pop II or even pop I stars. True pop III galaxies may, however, form at slightly later epochs, in low-density environments which have remained chemically pristine (e.g. \cite{Stiavelli & Trenti,Johnson,Trenti et al.}). Since pockets of primordial gas may survive in galaxies that have already experienced some chemical enrichment, hybrid galaxies in which pop III, II and I stars  continue to form in parallel can also be expected (e.g. \cite{Salvaterra et al.}). Even more exotic galaxies may be envisioned if the dark matter of the Universe has the properties required for the formation of long-lived population III stars fueled by WIMP annihilations in minihalos (e.g. \cite{Spolyar et al.}). Many such ``dark stars'' may then end up within the first galaxies during their hierarchial assembly, and possibly imprint detectable signatures in their spectra \cite{Zackrisson et al. a}. 

The James Webb Space Telescope\footnote{http://www.jwst.nasa.gov/} (JWST), scheduled for launch in 2014, may in principle allow the first direct detections of massive population III stars, but this is no doubt going to be very challenging -- both because of the faintness of these sources (pop III galaxies; conventional pop III stars or dark stars) at the relevant redshifts, but also because of the difficulties in identifying these rare objects in the vast amounts data provided by planned JWST surveys. Here, we present {\it Yggdrasil} -- the first spectral synthesis model custom-designed for the first galaxies -- and use it to predict the smallest stellar population masses (for both pop I, II and III galaxies) detectable with JWST, as a function of redshift. We also explore the prospects of a proposed JWST survey called {\it Palantir}, which will exploit the gravitational magnification provided by foreground galaxy clusters to hunt for the first stars and galaxies at $z\gtrsim 10$.

\section{The spectral evolution of the first galaxies}
The {\it Yggdrasil}\footnote{named after a sacred tree in Norse mythology} model \cite{Zackrisson et al. b} is a population synthesis model custom-designed for modelling the spectral energy distributions (SEDs) of these first galaxies. To reflect the significant variance in terms of stellar content that these objects may display, {\it Yggdrasil} is equipped to handle mixtures of conventional pop I, II and III stars as well as dark stars. It also includes nebular emission from the photoionized gas and extinction due to dust. A number of {\it Yggdrasil} model results are already publicly available from the author's homepage\footnote{ www.astro.su.se/$\sim$ez}, and more will soon be added. 

The SEDs of Single Stellar Populations (SSPs\footnote{also known as single-age stellar populations or instantaneous-burst populations}), from various other population synthesis models can be used as input to {\it Yggdrasil}, which then reweights the SSP time steps to accommodate arbitrary star formation histories. For the duration of this paper, we will adopt Starburst99 (\cite{Leitherer et al.,Vazquez & Leitherer}) SSPs generated with Padova-AGB stellar evolutionary tracks and the Kroupa \cite{Kroupa} universal stellar initial mass function (IMF) throughout the mass range 0.1--100 $M_\odot$ for population II (assumed metallicity $Z=0.0004$) and pop I (assumed metallicity $Z=0.020$) galaxies.

While both theoretical arguments and numerical simulations support the notion that pop III stars must have been more massive than the pop II and I stars forming later on (for a review, see \cite{Bromm & Larson}), the exact IMF of pop III stars remains unknown. It has been argued that the Universe may have produced two classes of pop III stars -- pop III.1 stars which formed first, with characteristic masses around $\sim 100\ M_\odot$, and pop III.2 stars which formed somewhat later and had lower characteristic masses of ($\sim 10\ M_\odot$) due to HD cooling promoted by the Lyman-Werner feedback provided by the pop III.1 stars (e.g. \cite{Greif & Bromm,Mackey et al.}). Even though the latest simulations suggest that the actual situation may be far more complicated (e.g. \cite{Stacy et al.,Clark et al.}, we have chosen to adopt this pop III.1/pop III.2 convention throughout this paper. 

Naively, one would expect the pop III.1 IMF to be appropriate for the stars forming in isolation (or in small numbers) in $\sim 10^{5-6}\ M_\odot$ minihalos capable of H$_2$ cooling, whereas the pop III.2 IMF may be more relevant for the first pop III galaxies forming in $\sim 10^{7-8}\ M_\odot$ halos capable of HI cooling. However, given the large uncertainties still attached to this picture, we here consider both pop III.1 and pop III.2 IMFs as plausible alternatives for pop III galaxies. For pop III.1 galaxies, we adopt the Schaerer stellar SSP \cite{Schaerer b} with a power-law IMF ($\mathrm{d}N/d\mathrm{M}\propto M^{-\alpha}$) with slope $\alpha=2.35$ throughout the mass range 50--500$ M_\odot$. For pop III.2 galaxies, we adopt the Raiter et al. TA model \cite{Raiter et al.}, which has a log-normal IMF with characteristic mass $M_\mathrm{c}=10 M_\odot$ and dispersion $\sigma=1\ M_\odot$. 

The contribution to the SED from photoionized gas is computed using the procedure outlined by Zackrisson et al. \cite{Zackrisson et al. c}. In this machinery, the stellar population SED is, at every age step, used as input to the photoionization code Cloudy \cite{Ferland et al.}. This results in a self-consistent prediction for the nebular continuum and emission line fluxes which reflects the temporal changes in the ionizing radiation. Throughout this paper, we will for simplicity assume that all of the ionizing radiation produced by stars within the model galaxies is absorbed locally and that no Lyman continuum radiation is escaping into the intergalactic medium, even though simulations of pop III galaxies suggest that substantial leakage may well occur \cite{Johnson et al. b}.
 
\begin{figure}
\centering
\includegraphics[width=.4\textwidth]{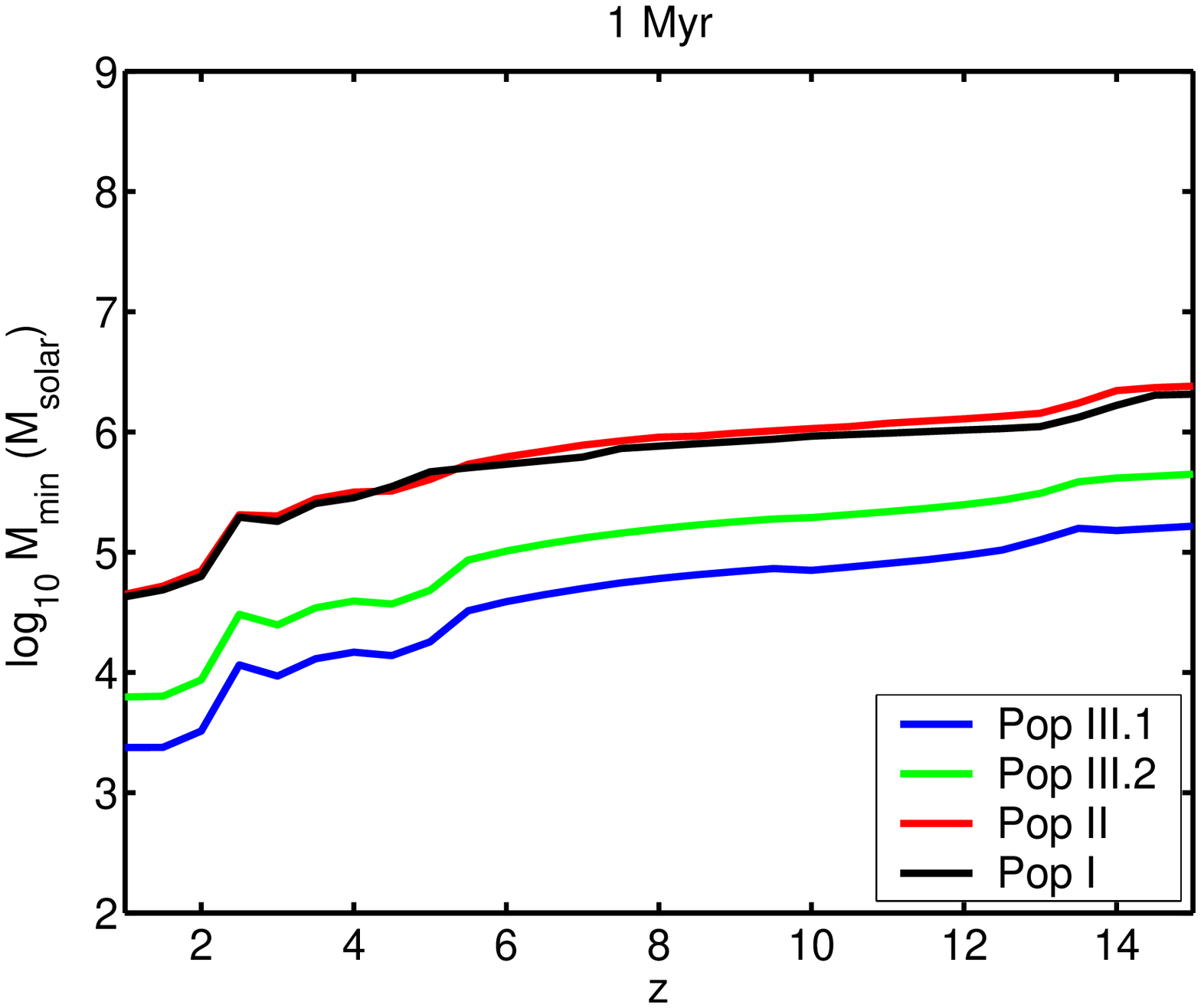}\includegraphics[width=.4\textwidth]{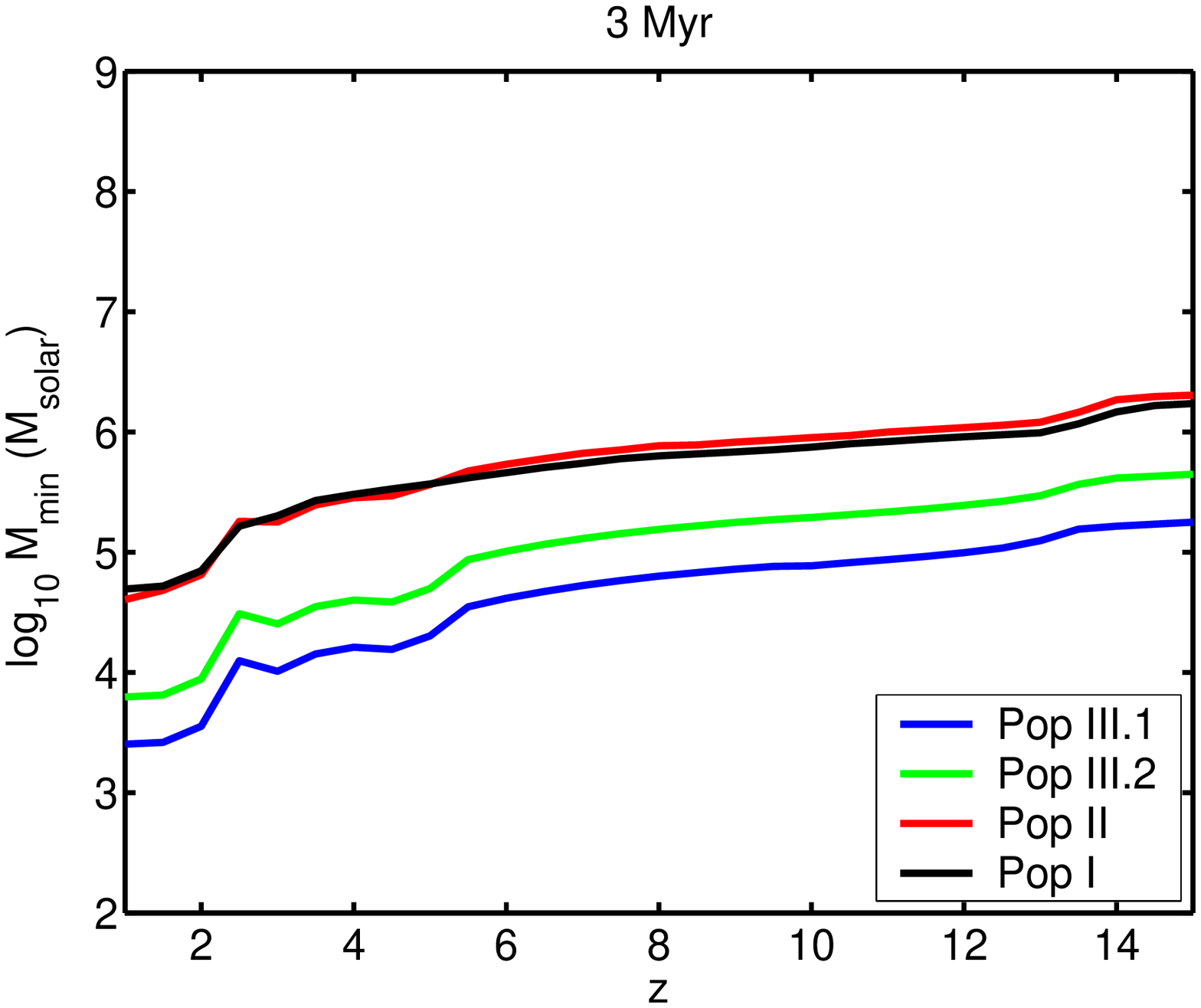}\\
\includegraphics[width=.4\textwidth]{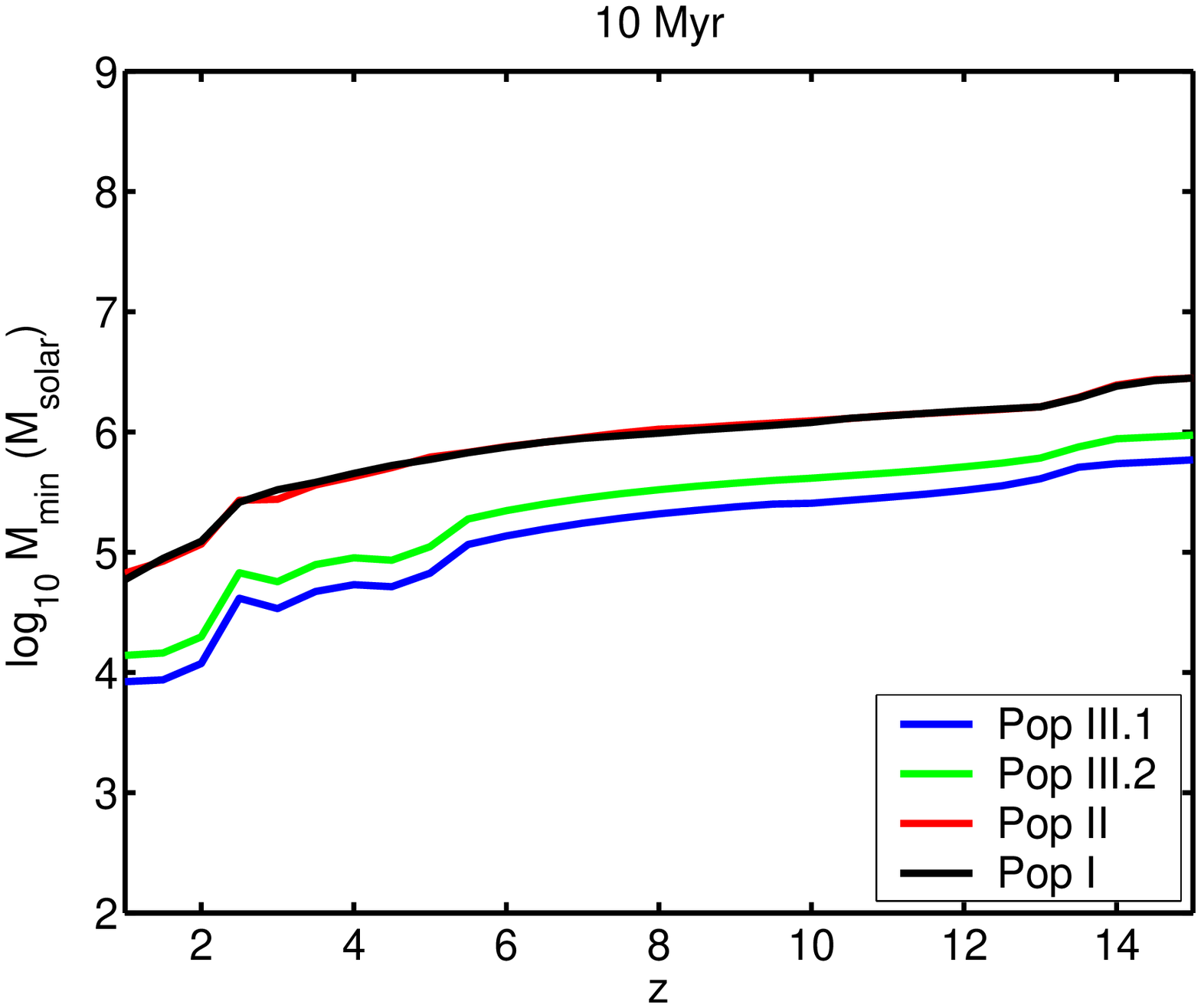}\includegraphics[width=.4\textwidth]{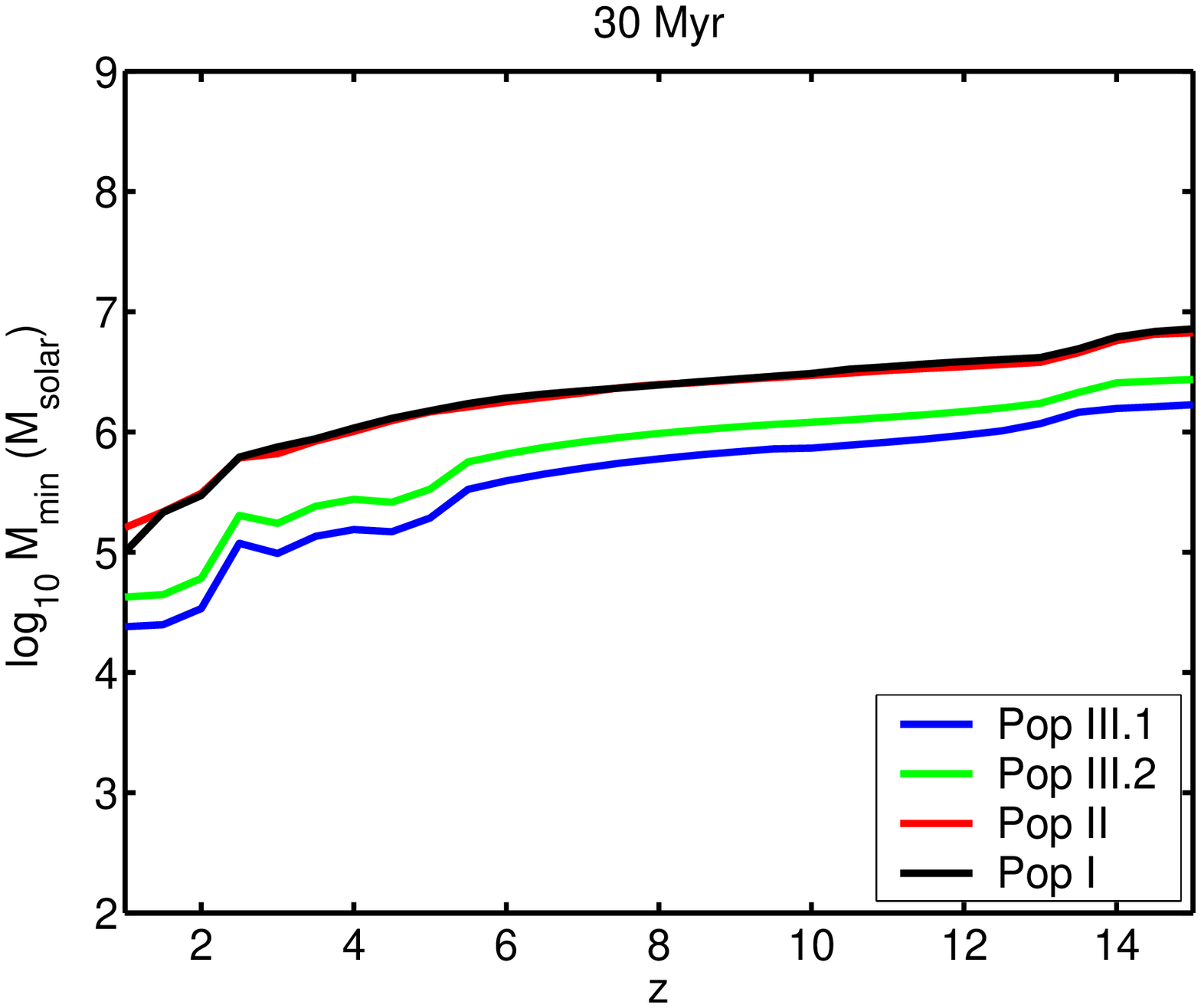}
\caption{Predicted mass detection limits in the JWST ultra deep field. The panels show the lowest masses of burst-like stellar populations that JWST may detect through broadband imaging (in filters with spectral resolution $R=4$) at 5$\sigma$ after a 100 h exposure, as a function of redshift. The line colours represent the different population metallicities and IMFs ($Z=0$ for pop III.1 and pop III.2, $Z=0.0004$ for pop II and $Z=0.020$ for pop I) -- see main text for details. Each panel corresponds to a different starburst age (1, 3, 10 and 30 Myr) for a population forming stars at a constant rate since $t=0$ yr.}
\label{fig1}
\end{figure}

In Fig.~\ref{fig1}, we use this model to predict the population masses of the faintest star-forming objects detectable though JWST broadband imaging in an ultra deep field (UDF), as a function of age and metallicity/IMF (pop III.1, pop III.2, pop II and pop I). These limits are based on the requirement that galaxies are detected at 5$\sigma$ in {\it at least} one JWST broadband filter (spectral resolution $R=4$) after a 100 h exposure (per filter). To avoid predictions hinging on the highly uncertain luminosity of the Ly$\alpha$ emission line at high redshifts, we assume the Ly$\alpha$ escape fraction to be zero -- i.e. the Ly$\alpha$ line does not contribute to the predicted fluxes at all. When computing the JWST broadband fluxes, we furthermore set all SED fluxes shortward of Ly$\alpha$ to zero for model galaxies at $z>6$, to reflect the high level of absorption in the neutral intergalactic medium at these epochs.

At $z<15$, and for the model spectra used here, these mass limits are typically determined by the predicted fluxes in the NIRCam F200W filter (the $R=4$ filter with the best sensitivity). Please note that we here use the mass of gas converted into stars since the beginning of a star formation episode. This is equivalent to the population mass often discussed in the context of simulations, where the overall gas mass is multiplied by the star formation efficiency of the first starburst episode to compute the gas converted into stars. The mass in {\it luminous stars} at these ages can be considerably lower at ages $\gtrsim 3$ Myr (especially in the case of pop III.1 and pop III.2), since many of the stars forming at $t=0$ yr have then already faded away. 

As seen in Fig.~\ref{fig1}, $M\sim10^6$--$10^7\ M_\odot$ star-forming pop I and pop II galaxies (black and red lines; mostly overlapping) can be detected at $z\approx 10$, whereas pop III.1 galaxies (blue lines) can be detected even if they have masses an order of magnitude lower. Pop III.2 galaxies (green lines) are intermediate between pop III.1 and pop I/II in these diagrams. The reason for the lower detection masses for pop III galaxies is primarily because of the higher relative contribution from nebular emission to the fluxes of these objects. The differences between the types of galaxies would not be as conspicuous if a substantial amount of Lyman continuum radiation were to escape into the intergalactic medium (as predicted in \cite{Johnson et al. b})

\section{Hunting for Pop III galaxies in lensed fields}
Unfortunately, some of the most exciting objects that might be hiding at high redshifts are likely to be too faint for JWST in UDF-style observations. This includes isolated pop III stars \cite{Gardner et al.,Greif et al.,Rydberg et al.}, dark stars with masses $M\lesssim 10^3\ M_\odot$ \cite{Zackrisson et al. a} and possibly also pop III galaxies. For example, the most massive pop III galaxies at $z\approx 10$ in the Trenti et al. \cite{Trenti et al.} simulations have baryonic masses of $\sim 10^7\ M_\odot$. According to Fig.~\ref{fig1}, at least $10^5\ M_\odot$ of these need to be converted into stars in the first star formation episode (i.e. star formation efficiency $\epsilon \gtrsim 10^{-2})$ in order for such an object to be detectable in a JWST UDF. The gas must also be sufficiently dense to keep the HII region confined inside the virial radius of the its host halo, otherwise ionizing radiation would be leaking into the intergalactic medium, with a lower overall luminosity as the result \cite{Johnson et al. b}. Fainter objects may, however, be detectable if one exploits the gravitational lensing provided by massive foreground objects. Galaxy clusters at $z\approx 0.1$--0.6 can in principle boost the fluxes of high-redshift objects by factors of $\mu\sim 10$--100 (e.g. \cite{Bradac et al.,Maizy et al.,Zitrin et al. a,Zitrin et al. b}). This means that one can reach significantly deeper than the JWST UDF, in just a fraction of the exposure time, by targeting a lensed field. The obvious drawback is that the high-redshift volume probed at the same time drops by a factor equal to the magnification $\mu$, which means that rare types of objects may be impossible to detect this way, even if lensing lifts their apparent magnitudes above the flux detection limit.

Our team is currently investigating the prospects of a proposed JWST survey called {\it Palantir}\footnote{named after a magical object in Lord of the Rings}, which will hunt for exotic high-redshift objects in fields lensed by foreground galaxy clusters. The primary target of the survey is the galaxy cluster MACS J0717.5+3745 at $z=0.546$ -- arguably the best lensing cluster currently available for studies of high-redshift objects \cite{Zitrin et al. a} -- but other clusters with similar properties (e.g. \cite{Zitrin et al. b}) may possibly also be covered. In \cite{Zackrisson et al. a}, we demonstrate that $M\lesssim 10^3\ M_\odot$ dark stars may in principle be within reach of a {\it Palantir}-style survey, but this requires that a very high fraction ($\gtrsim 0.1$) of the minihalos harbouring pop III stars produce long-lived dark stars (with lifetimes $\gtrsim 10^7\ M_\odot$), and that very long JWST exposures ($\approx 30$ h per filter) are used. The prospects of detecting pop III galaxies may, however, be substantially brighter.

By projecting the pop III galaxy simulations by Trenti et al. \cite{Trenti et al.} through the MACS J0717.5+3745 magnification maps, and using {\it Yggdrasil} to derive the intrinsic fluxes of these objects, we predict that {\it Palantir} should be able to detect at least a handful of $z\approx 7-15$ pop III galaxies in about 3 hours of JWST NIRCam/F200W imaging, as long as the typical star formation efficiencies of these objects are $\epsilon \gtrsim 3\times 10^{-3}$. Here, we have assumed an instantaneous-burst population with a pop III.2 IMF (model TA from \cite{Raiter et al.}; lognormal IMF centered on 10 $M_\odot$) and no Lyman continuum leakage. We moreover impose an upper age limit of 5 Myr, after which we assume that these galaxies lose their spectral identities as pop III galaxies due to subsequent, metal-enriched star formation. By comparison, a $z=10$ galaxy of this type would need to be a factor of $\approx5$ more massive to be detectable in a JWST UDF. Hence, there is a good chance of detecting objects that are intrinsically fainter than the JWST UDF detection limit, in just a fraction of the observing time (3\% in this case).  

\section{Summary}
We present a new spectral synthesis code designed to predict the SEDs of the first galaxies (pop I, II and III with or without nebular emission, dust extinction and dark stars). Using this model, we find that Pop III galaxies may be sufficiently bright to be detectable at $z=10$ through JWST ultra-deep imaging (100 h exposures on a single, unlensed field), but this requires low Lyman continnum escape fractions ($f_\mathrm{esc}\sim 0$) and relatively high star formation efficiencies ($\epsilon\gtrsim 0.01$). Pop III galaxies that are at least factor of $\approx5$ times less luminous may, however, be detectable in lensed fields in just a fraction ($\approx 3$\%) of the observing time. We propose a JWST survey called {\it Palantir} which will hunt for faint, exotic objects (pop III galaxies, dark stars etc.) behind galaxy clusters like MACS J0717.5+3745 -- arguably the best lensing cluster currently known for surveys of this type.  

\section*{Acknowledgments}
Daniel Schaerer, Michele Trenti, Adi Zitrin, Tom Broadhurst, Claes-Erik Rydberg and G\"oran \"Ostlin are acknowledged for fruitful collaboration on the topic of cluster lensing and population III galaxies.

\end{document}